\documentclass[preprint,prb]{revtex4}
\usepackage{amsmath}

\usepackage{amssymb}
\usepackage{amsthm}

\newcommand{{\bff}}{\mbox{\boldmath$f$\unboldmath}}

\newcommand{{\bfF}}{\mbox{\boldmath$F$\unboldmath}}

\newcommand{{\bfA}}{\mbox{\boldmath$A$\unboldmath}}

\def\v#1{{\bf#1}}
\begin{document}

\title{How the potentials in different gauges yield the same retarded electric and magnetic fields}
\author{Jos\'e A. Heras}
\email{heras@phys.lsu.edu}
\affiliation{Departamento de F\'\i sica, E. S. F. M., Instituto
Polit\'ecnico Nacional, M\'exico D. F. M\'exico and 
Department of Physics and Astronomy, Louisiana State University, Baton
Rouge, Louisiana 70803-4001, USA}
\begin{abstract}
This paper presents a simple and systematic method to show how the potentials in the Lorentz, Coulomb, Kirchhoff, velocity and temporal gauges yield the same retarded electric and magnetic fields. The method appropriately uses the dynamical equations for the scalar and vector potentials 
to obtain two wave equations, whose retarded solutions lead to the electric and magnetic fields. The advantage of this method is that it does not use explicit expressions for the potentials in the above gauges, which are generally simple to obtain for the scalar potential but generally difficult to calculate for the vector potential. The spurious character of the term generated by the scalar potential in the Coulomb, Kirchhoff and velocity gauges is noted.  The non spurious character of the term generated by the scalar potential in the Lorenz gauge is emphasized.
\end{abstract}
\maketitle
\section{Introduction}
As is well-known in electromagnetism, the {\it advantage} of the Coulomb gauge is that the scalar potential in this gauge is particularly simple to obtain but the {\it disadvantage} is that the vector potential in this gauge is particularly difficult to calculate. This characteristic of the Coulomb gauge can explain why the explicit demonstration that the potentials in this gauge yield the retarded electric and magnetic fields is not usually presented in textbooks. In the best of cases, some of these textbooks\cite{1,2} mention a paper of Brill and Goodman\cite{3} in which an elaborate proof  that the potentials in the Coulomb and Lorenz gauges yield the same retarded electric and magnetic fields is presented ---this proof is restricted to sources with harmonic time dependence. 

The explicit proof that the Coulomb-gauge potentials yield the retarded electric and magnetic fields is conceptually important to emphasize the fact that the causal behavior of these fields is never effectively lost when they are expressed in terms of such potentials, despite the result that the scalar potential in this gauge propagates instantaneously, which would seem to indicate a {\it lost} of causality in the electric field. Similar proofs that potentials in other gauges like the temporal or velocity gauges yield the retarded electric and magnetic fields are omitted in textbooks. 
This omission is, however, comprehensible because these other gauges are not usually mentioned in textbooks of electrodynamics. Actually, the velocity and temporal gauges are less-known than the Coulomb or Lorenz gauges. The velocity gauge is one in which the scalar potential propagates with an arbitrary velocity and the temporal gauge is one in which the scalar potential is identically zero.

In a recent paper Jackson and Okun\cite{4} have reviewed the interesting history that led to the conclusion that potentials in different gauges describe the same physical fields. In a subsequent paper, Jackson\cite{5} derived novel expressions for the vector potential in the Coulomb, velocity and temporal gauges and demonstrated explicitly how these expressions for the vector potential together with their associated expressions for the scalar potential originate the same retarded electric and magnetic fields.  Jackson emphasized:\cite{5} ``... whatever propagation or nonpropagation characteristics are exhibited by the potentials in a particular gauge, the electric and magnetic fields are always the same and display the experimentally verified properties of causality and propagation at the speed of light." 
Rohrlich\cite{6}  has also recently discussed causality in the Coulomb gauge. The present author has used two different methods \cite{7,8} to show that the Coulomb-gauge potentials yield the retarded electric field and in a subsequent paper\cite{9} has re-discovered the Kirchhoff gauge,\cite{10} in which the scalar potential ``propagates" with the imaginary speed $``ic"$, where $c$ is the speed of light. In a recent paper,
Yang\cite{11} has discussed again the velocity gauge.\cite{12} 
 
To show that potentials in different gauges yield the same retarded fields, one usually first derives explicit expressions for the scalar and vector potentials in a specified gauge. The retarded fields are then obtained by differentiation of such expressions for potentials. The practical difficulty of this usual method is that the derivation of explicit expressions for potentials in most gauges is not generally a simple task, particularly for the case of the vector potential. The question then arises: Is it necessary to have explicit expressions for potentials in a specified gauge to show that they lead to the retarded electric and magnetic fields? The answer is {\it negative}, at least for 
the gauges considered in this paper.

In this paper we present a simple and systematic method to show how the potentials in the Lorentz, Coulomb, Kirchhoff, velocity and temporal gauges yield the same retarded electric and magnetic fields. Instead of using explicit expressions for the scalar and vector potentials in the above gauges, we appropriately use the dynamical equations of potentials to obtain two wave equations, whose retarded solutions lead to the retarded fields. An advantage of the proposed method is that it allows one to identify the spurious character of the gradient of the scalar potential in the Coulomb, Kirchhoff and velocity gauges. We emphasize the non spurious character of the gradient of the Lorenz-gauge scalar potential. Finally, we suggest that the  Lorenz-gauge potentials could be interpreted as physical objects.

In Sec.~II we define the Lorentz, Coulomb, Kirchhoff, velocity and temporal gauges. In Sec.~III we review the usual proof that the Lorentz-gauge potentials lead to the retarded fields and apply the alternative method to show how these potentials yield the retarded fields. 
In Sec.~IV we apply the method to the  Coulomb-gauge potentials. 
In Sec.~V we clearly define the steps of the proposed method. In Sec.~VI we apply the method to the Kirchhoff-gauge potentials. In Sec.~VII we apply the method to the velocity-gauge potentials. In Sec.~VIII we apply method to the temporal-gauge vector potential. In Sec.~IX we emphasize the spurious character of the gradient of the scalar potential in the Coulomb, Kirchhoff and velocity gauges as well as the non spurious character of the gradient of the scalar potential in the Lorenz gauge. We suggest that the Lorenz-gauge potentials may be interpreted as physical quantities. In Sec.~X we present some concluding remarks.

\section{electromagnetic gauges}
It is well-known that the electric and magnetic fields $\v E$ and $\v B$ are determined from the scalar and vector potentials $\Phi$ and $\v A$ according to
\begin{subequations}
\begin{align}
\v E& =\!-\nabla\Phi- \frac{\partial\v A}{\partial t},\\
\v B & =\!\nabla\times\v A.
\end{align}
\end{subequations}
Here we are using SI units and considering fields with localized sources in vacuum.
The fields $\v E$ and $\v B$ are shown to be invariant under the gauge transformations 
 \begin{subequations}
\begin{align}
\Phi'&=\!\Phi-\frac{\partial \chi}{\partial t},\\
\v A'&= \!\v A +\nabla\chi,
\end{align}
\end{subequations}
where $\chi$ is an arbitrary time-dependent gauge function. The inhomogeneous Maxwell equations together with Eqs.~(1) lead to the coupled equations
\begin{subequations}
\begin{align}
\nabla^2\Phi &=\! -\frac{\rho}{\epsilon_0} -\frac{\partial}{\partial t}(\nabla\cdot\v A),\\
\Box^2\v A&=\!-\mu_0\v J + \nabla\left(\nabla\cdot\v A+\frac{1}{c^2}\frac
{\partial \Phi}{\partial t}\right),
\end{align}
\end{subequations}
where $\Box^2\equiv\nabla^2-(1/c^2)\partial^2/\partial t^2$ is the D'Alambertian operator and $\rho$ and $\v J$ are the charge and current densities respectively. The arbitrariness of the gauge function $\chi$ in Eqs.~(2) allows one to choose a gauge condition. We will consider here five gauge conditions, 
\vskip 5pt
\noindent Lorenz gauge:\cite{1,2}
\begin{equation}
\nabla\cdot\v A +\frac{1}{c^2}\frac{\partial\Phi}{\partial t}=0.
\end{equation} 
Coulomb gauge:\cite{3,5}
\begin{equation}
\nabla\cdot\v A=0.
\end{equation}
Kirchhoff gauge:\cite{9}
\begin{equation}
\nabla\cdot\v A -\frac{1}{c^2}\frac{\partial\Phi}{\partial t}=0.
\end{equation}
Velocity gauge:\cite{5,11,12}
\begin{equation}
\nabla\cdot\v A + \frac{1}{v^2}\frac{\partial\Phi}{\partial t}=0.
\end{equation}
Temporal gauge:\cite{5}
\begin{equation}
\Phi=0.
\end{equation}
We note that the velocity gauge is a class of gauges that contains the first three gauges: the Lorenz gauge $(v=c)$,  the Coulomb gauge $(v=\infty)$, and
the Kirchhoff gauge $(v=ic)$. Most textbooks discuss the Lorenz gauge and briefly mention the Coulomb gauge.\cite{1,2} In contrast, the velocity and temporal gauges are not usually mentioned in textbooks. The Kirchhoff gauge\cite{10} is not also mentioned, which is comprehensible because this gauge has recently been re-discovered.\cite{9} There are other gauges like the non-relativistic and relativistic multipolar gauges but they will be not considered here.\cite{5}  

\section{LORENZ GAUGE}
The most popular gauge is the Lorenz gauge which allows one to uncouple Eqs.~(3) in such a way that the scalar and vector potentials 
are described by symmetrical (uncoupled) equations, which is a peculiar characteristic of this gauge. In fact, if we assume the Lorenz gauge (4) then Eqs.~(3) become the wave equations:
\begin{subequations}
\begin{align}
\square^2\Phi_L&=\! -\frac{\rho}{\epsilon_0},\\
\square^2\v A_L&=\!-\mu_0\v J.
\end{align}
\end{subequations}
The notation $\Phi_L$ and $\v A_L$ indicates that these potentials are in the Lorenz gauge. An advantage of the Lorenz gauge is that the retarded solutions of Eqs.~(9) are well-known,
\begin{subequations}
\begin{align}
\Phi_L(\v x,t)&=\!\frac{1}{4\pi\epsilon_0}\int d^3x'\frac{1}{R}\rho(\v x',t-R/c),\\
\v A_L(\v x,t)&=\!\frac{\mu_0}{4\pi}\int d^3x'\frac{1}{R}\v J(\v x',t-R/c),
\end{align}
\end{subequations}
where $R= |\v x-\v x'|$ is the magnitude of the vector $\v R= \v x-\v x'$ with $\v x$ the field point and $\v x'$ the source point. The integrals in Eqs.~(10) are extended over all space. 
An advantage of the Lorenz gauge condition is that it can be written in a relativistically covariant form: $\partial_\mu A^\mu=0,$ where  $\partial_\mu\equiv\left((1/c)\partial/\partial t,\nabla\right)$ and $A^\mu\equiv\left(\Phi/c,\v A\right)$. Greek indices run from 0 to 3; the signature of the Minkowski metric is (1,-1,-1,-1) and summation on repeated indices is understood. Equations (9) an (10) can also be written in a relativistically covariant form. Another characteristic of the Lorenz gauge is that the scalar potential in this gauge yields a retarded term which can be written as
\begin{equation}
-\nabla \Phi_L(\v x,t)=\frac{1}{4\pi\epsilon_0}\int d^3 x'\left(\frac{\hat{\v R}}{R^2}\rho(\v x',t-R/c)+\frac{\hat{\v R}}{Rc}\frac{\partial\rho(\v x',t-R/c)}{\partial (t-R/c)}\right),
\end{equation}
where $\hat{\v R}=\v R/R$. This term displays the experimentally verified properties of causality and propagation at the speed of light. We anticipate that the gradient of the scalar potential in the other gauges considered in this paper does not generally satisfy the above properties.

Having obtained the potentials $\Phi_L$ and $\v A_L$, the electric and magnetic can be derived by the usual prescription 
\begin{subequations}
\begin{align}
\v E&=\!-\nabla\Phi_L-\frac{\partial \v A_L}{\partial t}=-\nabla\left(\frac{1}{4\pi\epsilon_0}\int d^3x'\frac{1}{R}[\rho]\right)-\frac{\partial}{\partial t}\left(\frac{\mu_0}{4\pi}\int d^3x'\frac{1}{R}[\v J]\right),\\
\v B&=\!\nabla\times\v A_L=\nabla\times\left(\frac{\mu_0}{4\pi}\int d^3x'\frac{1}{R}[\v J]\right),
\end{align}
\end{subequations}
where we have introduced the retardation symbol $[\quad]$ to indicate that the enclosed quantity is to be evaluated at the retarded time $t'=t-R/c$. Henceforth, all square brackets will indicate retardation.

After an integration by parts, Eqs.~(12) become the usual electric and magnetic fields
\begin{subequations}
\begin{align}
\v E &=\!\frac{1}{4\pi\epsilon_0}\int d^3x'\frac{1}{R}\left[ -\nabla'\rho-\frac{1}{c^2}\frac{\partial \v J}{\partial t' }\right],\\
\v B &=\!\frac{\mu_0}{4\pi}\int d^3x'\frac{1}{R}[\nabla'\times\v J].
\end{align}
\end{subequations}
The result expressed in Eqs.~(13) is well-known. However, Eqs.~(13) can also be obtained without considering Eqs.~(10) by applying an alternative method. We first take  minus the gradient to Eq.~(9a) and minus the time derivative to Eq.~(9b),
\begin{subequations}
\begin{align}
-\Box^2\nabla\Phi_L&=\!\frac{\nabla\rho}{\epsilon_0},\\
-\Box^2\frac{\partial \v A_L}{\partial t }&=\!\mu_0\frac{\partial \v J}{\partial t }.
\end{align}
\end{subequations}
The retarded solutions of Eqs.~(14) are given by 
\begin{subequations}
\begin{align}
-\nabla\Phi_L &=\!\frac{1}{4\pi\epsilon_0}\int d^3x'\frac{1}{R}\left[ -\nabla'\rho\right],\\
-\frac{\partial \v A_L}{\partial t }&=\!\frac{\mu_0}{4\pi}\int d^3x'\frac{1}{R}\left[-\frac{\partial \v J}{\partial t' }\right].
\end{align}
\end{subequations}
On the other hand, the curl of Eq.~(9b) gives
\begin{equation}
\square^2\left(\nabla\times\v A_L\right)=-\mu_0\nabla\times\v J,
\end{equation}
with the retarded solution 
\begin{equation}
\nabla\times\v A_L=\frac{\mu_0}{4\pi}\int d^3x'\frac{1}{R}[\nabla'\times\v J].
\end{equation}
Equations (15) and (17) yield the usual form of the retarded electric and magnetic fields
\begin{subequations}
\begin{align}
\v E&=\!-\nabla\Phi_L-\frac{\partial \v A_L}{\partial t}=\frac{1}{4\pi\epsilon_0}\int d^3x'\frac{1}{R}\left[ -\nabla'\rho-\frac{1}{c^2}\frac{\partial \v J}{\partial t' }\right],\\
\v B&=\!\nabla\times\v A_L=\frac{\mu_0}{4\pi}\int d^3x'\frac{1}{R}[\nabla'\times\v J].
\end{align}
\end{subequations}
This alternative method to obtain the electric and magnetic fields, which works directly with Eqs.~(9), seems not to have some practical advantage with respect to the traditional method that uses Eqs.~(10). This is so because we are considering potentials in the Lorenz gauge. But we will see in the next sections that this method will be advantageous when applied to potentials in other gauges.

\section{COULOMB GAUGE}

A less popular gauge in textbooks is the Coulomb gauge. In this gauge the scalar potential satisfies an instantaneous Poisson equation, which is a peculiar characteristic of this gauge. If we assume the Coulomb gauge (5) then Eqs.~(3) become the coupled equations:
\begin{subequations}
\begin{align}
\nabla^2\Phi_C&=\! -\frac{\rho}{\epsilon_0},\\
\square^2\v A_C&=\!-\mu_0\v J +\nabla\frac{1}{c^2}\frac
{\partial \Phi_C}{\partial t},
\end{align}
\end{subequations}
where we have used the notation $\Phi_C$ and $\v A_C$ to specify that these potentials are in the Coulomb gauge. As pointed out in the introduction,
the advantage of the Coulomb gauge is that the solution of Eq.~(19a) is particularly simple to obtain but the disadvantage is that the solution of 
Eq.~(19b) is particularly difficult to calculate. Let us write the solutions of Eqs.~(19) in an explicit form as follows
\begin{subequations}
\begin{align}
\Phi_C(\v x,t)&=\!\frac{1}{4\pi\epsilon_0}\int d^3 x'\frac{1}{R}\rho(\v x',t),\\
\v A_C(\v x,t)&=\!\frac{\mu_0}{4\pi}\int d^3 x'\frac{1}{R}\left(\left\{\v J(\v x',t-R/c)-c\hat{\v R}\rho(\v x',t-R/c)\right\} +\frac{c^2\hat{\v R}}{R} \int_{0}^{R/c}d\tau\rho(\v x',t-\tau)\right),
\end{align}
\end{subequations}
Equation (20a) is a well-known instantaneous expression and Eq.~(20b) is a novel expression derived recently by Jackson.\cite{5} By making use of Eqs.~(20), Jackson\cite{5} has obtained the retarded electric field in the form given by Jefimenko\cite{13} and the usual retarded form of the magnetic field given in Eq.~(13b). A disadvantage of the Coulomb gauge condition is that it cannot be written in a relativistically covariant form. 

The scalar potential in the Coulomb gauge yields the instantaneous term 
\begin{equation}
-\nabla\Phi_C(\v x,t)=\frac{1}{4\pi\epsilon_0}\int d^3 x'
\frac{\hat{\v R}}{R^2}\rho(\v x',t).
\end{equation}
This term does not displays the experimentally verified properties of causality and propagation at the speed of light and therefore 
its explicit presence in the expression for the retarded electric field $\v E = -\nabla\Phi_C-\partial \v A_C/\partial t$ seems to indicate 
at first sight an undesirable inconsistency. We recall that an instantaneous field is actually in conflict with special relativity which 
states that no physical information can propagate faster than $c$ in vacuum. 

In order to understand the role played by the acausal term $-\nabla\Phi_C$ 
in the electric field $\v E = -\nabla\Phi_C-\partial \v A_C/\partial t$, let us apply the method used in the previous section but now to  show that the potentials $\Phi_C$ and $\v A_C$ lead to the fields $\v E$ and $\v B$. 

In a {\it first} step we symmetrize Eq.~(19a) with respect to Eq.~(19b) by adding the term $-(1/c^2)\partial^2\Phi_C/\partial t^2$ on both sides of Eq.~(19a) to obtain the equation 
\begin{equation}
\square^2\Phi_C= -\frac{\rho}{\epsilon_0} -\frac{1}{c^2}\frac
{\partial^2 \Phi_C}{\partial t^2}. 
\end{equation}

In a {\it second} step, we take minus the gradient to Eq.~(22) and minus the time derivative to Eq.~(19b) to obtain two equations involving  
third-order derivatives of potentials,
\begin{subequations}
\begin{align}
-\square^2\nabla\Phi_C&=\!\frac{1}{\epsilon_0}\nabla\rho+\nabla\frac{1}{c^2}\frac
{\partial^2 \Phi_C}{\partial t^2},\\
-\square^2\frac{\partial \v A_C}{\partial t }&=\!\mu_0\frac{\partial \v J}{\partial t }-\nabla\frac{1}{c^2}\frac
{\partial^2 \Phi_C}{\partial t^2}.
\end{align}
\end{subequations}

In a {\it third} step we add these equations to obtain the wave equation 
\begin{equation}
\square^2 \left(- \nabla\Phi_C-\frac{\partial \v A_C}{\partial t }\right)= \frac{1}{\epsilon_0}\nabla\rho+\mu_0\frac{\partial \v J}{\partial t },
\end{equation}
with the retarded solution 
\begin{equation}
-\frac{\partial \v A_C}{\partial t }= \frac{1}{4\pi\epsilon_0}\int d^3x'\frac{1}{R}\left[ -\nabla'\rho-\frac{1}{c^2}\frac{\partial \v J}{\partial t' }\right] + \nabla\Phi_C.
\end{equation}
This expression states that the term $-\partial \v A_C/\partial t$ always contains the instantaneous component $\nabla\Phi_C$, which cancels exactly the instantaneous part $- \nabla\Phi_C$ of the electric field $\v E=-\nabla\Phi_C-\partial \v A_C/\partial t.$ This well-known result\cite{3} has recently been emphasized.\cite{5,14} However, the simple demonstration  of Eq.~(25) presented here is remarkable. Equation (25) has also recently been obtained by applying another more complicated method.\cite{15} Stated in other words: the explicit presence of an acausal term $(-\nabla\Phi_C)$ in the electric field expressed in terms of the Coulomb gauge potentials is {\it irrelevant} because such a term is always canceled by one of the components $(\nabla\Phi_C)$ of the remaining term defined by the Coulomb-gauge vector potential ($-\partial \v A_C/\partial t$). The field $-\nabla\Phi_C$ is then physically undetectable and can be interpreted as a $spurious$ field which exists mathematically but not physically. This means that causality is never effectively lost in the electric field. Accordingly, when Eq.~(25) is used into $\v E=-\nabla\Phi_C-\partial \v A_C/\partial t$ we obtain the usual retarded form of the electric field
\begin{equation}
\v E=-\nabla\Phi_C-\frac{\partial \v A_C}{\partial t}=\frac{1}{4\pi\epsilon_0}\int d^3x'\frac{1}{R}\left[ -\nabla'\rho-\frac{1}{c^2}\frac{\partial \v J}{\partial t' }\right],
\end{equation}

In a {\it fourth} step we take the curl to Eq.~(19b) to obtain the wave equation
\begin{equation}
\square^2\left(\nabla\times\v A_C\right)=-\mu_0\nabla\times\v J,
\end{equation}
with the retarded solution
\begin{equation}
\nabla\times\v A_C =\frac{\mu_0}{4\pi}\int d^3x'\frac{1}{R}[\nabla'\times\v J].
\end{equation}
Evidently, Eq. (28) is identified with the usual retarded form of the magnetic field
\begin{equation}
\v B=\nabla\times\v A_C =\frac{\mu_0}{4\pi}\int d^3x'\frac{1}{R}[\nabla'\times\v J].
\end{equation}
Neither the simple solution (20a) of Eq.~(19a) nor the complicated solution (20b) of Eq.~(19b) have been required to show that the Coulomb-gauge potentials yield the electric and magnetic fields. This is the main advantage of the method proposed here.

\section{THE FOUR STEPS OF THE METHOD}
It would be useful for the reader to have an explicit definition of the
steps of the method proposed in previous sections. As noted in Sec.~IV, the method can be defined by four steps:
\begin{description}

\item [Step 1.] Apply some of the Lorentz, Coulomb, Kirchhoff, velocity and temporal gauges to equations (3) and symmetrize, if necessary,  the {\it gauged} equation for the charge density with respect to the {\it gauged} equation for the current density. 

\item [Step 2.] Calculate minus the gradient of the gauged (and possibly symmetrized) equation for the charge density and minus the time derivative of the gauged equation for the current density. As a result, two equations containing third-order derivatives of potentials are obtained: one involving the gradient of the charge density and the other one involving the time derivative of the current density. 

\item [Step 3.] Except in the case of the Lorenz guage, to add the third-order equations obtained in the step 2 to obtain a wave equation, whose retarded solution gives an equation for the time derivative of the vector potential, which is substituted into the expression for the electric field in terms of potentials to obtain the retarded electric field.  

\item [Step 4.] Take the curl to the third-order equation for the current density obtained in the step 2 
to obtain a wave equation, whose retarded solution gives an equation for the curl of the vector potential, which is substituted into the expression for the magnetic field in terms of the vector potential to obtain the retarded magnetic field.  
\end{description}

In the exceptional case of the Lorentz gauge, it is necessary first to solve the equations derived in the step 2 [Eqs.~(14)]. The retarded solutions of these equations [Eqs.~(15)] are used in the step 3 to obtain the usual retarded form of the electric field.\cite{16}

To illustrate the steps of the four-steps method, we will consider  the cases of the Kirchhoff, velocity and temporal gauges in the next sections.

\section{KIRCHHOFF GAUGE}

According to the authors of Ref.~1, the first published relation between potentials is due to Kirchhoff\cite{10} who showed that the Weber form of the vector potential $\v A$ and its associated scalar potential $\Phi$ satisfy the equation (in modern notation): $\nabla\cdot\v A -1/(c^2)\partial\Phi/\partial t=0,$ that is, Eq.~(6), which was originally obtained for quasistatic potentials in which retardation is neglected. 
Of course, the electromagnetic gauge invariance had not been established yet in that time. The present author has proposed to call Eq.~(6) the Kirchhoff gauge and has presented a general discussion on this gauge.\cite{9}  
In the Kirchhoff gauge (6), Eqs.~(3) become 
\begin{subequations}
\begin{align}
\nabla^2\Phi_K+\frac{1}{c^2}\frac{\partial^2\Phi_K}{\partial t^2}&=\! -\frac{\rho }{\epsilon_0},\\
\square^2\v A_K &=\!-\mu_0\v J + \frac{2}{c^2}\nabla\frac
{\partial \Phi_K}{\partial t},
\end{align}
\end{subequations}
where we have used the notation $\Phi_K$ and $\v A_K$ to specify that these potentials are in the Kirchhoff gauge. 
We note that Eq.~(30a) is an elliptical equation, which does not describe a real propagation. Interestingly, Eq.~(30a) can be written as pseudo wave equation. In fact, after the simple substitution $c^2=-(ic)^2,$ Eq.~(30a) may be written as 
\begin{equation}
\nabla^2\Phi_K-\frac {1}{(ic)^2} \frac{\partial^2\Phi_K}{\partial t^2}= -\frac{\rho}{\epsilon_0}.
\end{equation}
This equation formally states that $\Phi_K$ ``propagates" with an imaginary speed $``ic"$ which emphasizes the unphysical character of $\Phi_K$.  The solutions of Eqs.~(30) can be expressed as\cite{9} 
\begin{subequations}
\begin{align}
\Phi_K(\v x,t)&=\!\frac{1}{4\pi\epsilon_0}\int d^3 x'\frac{1}{R}\rho(\v x',t-R/(ic))\\
\v A_K(\v x,t)&=\!\frac{\mu_0}{4\pi}\int d^3 x'\frac{1}{R}\bigg(\left\{\v J(\v x',t-R/c)-c\hat{\v R}\rho(\v x',t-R/c)\right\} +\frac{c}{i}\hat{\v R}\rho(\v x',t-R/(ic))\nonumber\\
&\quad+\frac{c^2\hat{\v R}}{R} \int_{R/(ic)}^{R/c}d\tau\;\rho(\v x',t-\tau)\bigg).
\end{align}
\end{subequations}
It has been noted in Ref.~9 that the potential $\Phi_K$ in Eq.~(32a) exhibits the same form that the scalar potential in the corresponding Lorenz gauge of an electromagnetic theory formulated in an Euclidean four-space.\cite{17,18} This interesting result shows how a same potential can be defined in {\it different gauges} and in {\it different spacetimes.} It is clear that the Kirchhoff gauge cannot be written in a relativistically covariant form. 
Also, in Ref.~9 it has been shown that Eqs.~(32) yield the retarded electric and magnetic fields.
However,  as shown in Ref.~9, the derivation of Eqs.~(32) is not so simple. 

We note that the potential $\Phi_K$ yields the imaginary term\cite{9} 
\begin{equation}
-\nabla \Phi_K(\v x,t)=\frac{1}{4\pi\epsilon_0}\int d^3 x'\left(\frac{\hat{\v R}}{R^2}\rho(\v x',t-R/(ic))+\frac{\hat{\v R}}{Ric}\frac{\partial\rho(\v x',t-R/(ic)}{\partial( t-R/(ic))}\right).
\end{equation}
The reader could find surprising the presence of the instantaneous term $-\nabla \Phi_C$ into the retarded expression of the electric field expressed in terms of the Coulomb-gauge potentials $\v E=-\nabla\Phi_C-\partial \v A_C/\partial t$.  But he probably would find even more surprising the presence of the {\it imaginary} term $-\nabla \Phi_K$ in the {\it observable} electric field expressed in terms of the Kirchhoff-gauge potentials $\v E=-\nabla\Phi_K-\partial \v A_K/\partial t.$ But the reader can reasonably suspect that, like the instantaneous term $-\nabla \Phi_C$, the imaginary term  $-\nabla \Phi_K$ does not play a physical role in the electric field. 
To understand the role of $-\nabla \Phi_K$ in Eq.~(33), let us to apply the method of the four steps proposed in Sec. V. We will show that the Kirchhoff potentials $\Phi_K$ and $\v A_K$ lead to the retarded fields $\v E$ and $\v B.$ 

Step 1. After applying the Kirchhoff condition (6) to Eqs.~(3), we have already obtained Eqs.~(30). We then $symmetrize$ Eq.~(30a) with respect to Eq.~(30b) by adding the term $-(2/c^2)\partial^2\Phi_K/\partial t^2$ on both sides of  Eq.~(30a) to obtain the equation 
\begin{equation}
\Box^2\Phi_K= -\frac{\rho }{\epsilon_0} -\frac{2}{c^2}\frac
{\partial^2 \Phi_K}{\partial t^2}. 
\end{equation}

Step 2. We take minus the gradient to Eq.~(34) and minus the time derivative to Eq.~(30b) to obtain the equations
\begin{subequations}
\begin{align}
-\Box^2\nabla\Phi_K&=\!\frac{\nabla\rho}{\epsilon_0}+\frac{2}{c^2}\nabla\frac
{\partial^2 \Phi_K}{\partial t^2},\\
-\Box^2\frac{\partial \v A_K}{\partial t }&=\!\mu_0\frac{\partial \v J}{\partial t }-\frac{2}{c^2}\nabla\frac
{\partial^2 \Phi_K}{\partial t^2}.
\end{align}
\end{subequations}

Step 3. We add Eqs.~(35) to obtain the wave equation 
\begin{equation}
\Box^2\bigg(- \nabla\Phi_K-\frac{\partial \v A_K}{\partial t }\bigg)= \frac{\nabla\rho}{\epsilon_0}+\mu_0\frac{\partial \v J}{\partial t },
\end{equation}
with the retarded solution 
\begin{equation}
-\frac{\partial \v A_K}{\partial t} =\frac{1}{4\pi\epsilon_0}\int d^3x'\frac{1}{R}\Bigg[ -\nabla'\rho-\frac{1}{c^2}\frac{\partial \v J}{\partial t' }\Bigg]+ \nabla\Phi_K.
\end{equation}
As may be seen, the term $-\partial \v A_K/\partial t$ in Eq.~(37) contains the imaginary component $\nabla\Phi_K$ which cancels exactly the imaginary part $- \nabla\Phi_K$ of the electric field $\v E=-\nabla\Phi_K-\partial \v A_K/\partial t.$ In other words: the explicit presence of an imaginary term $(-\nabla\Phi_K)$ in the electric field expressed in terms of the Kirchhoff-gauge potentials is irrelevant because such a term is always canceled by one of the components $(\nabla\Phi_K)$ of the Kirchhoff-gauge vector potential ($-\partial \v A_K/\partial t$). This result has recently been emphasized.\cite{14} We can state that the term $-\nabla\Phi_K$ is a $spurious$ field, which exists mathematically but not physically and consequently the {\it real} character of the electric field is never effectively lost. Accordingly, when Eq.~(37) is used in the field $\v E=-\nabla\Phi_K-\partial \v A_K/\partial t$, we obtain the usual retarded form of this field,
\begin{equation}
\v E=-\nabla\Phi_K-\frac{\partial \v A_K}{\partial t}=\frac{1}{4\pi\epsilon_0}\int d^3x'\frac{1}{R}\left[ -\nabla'\rho-\frac{1}{c^2}\frac{\partial \v J}{\partial t' }\right],
\end{equation}

Step 4. We take the curl to Eq.~(30b) to obtain the wave equation
\begin{equation}
\square^2\left(\nabla\times\v A_K\right)=-\mu_0\nabla\times\v J.
\end{equation}
with the retarded solution
\begin{equation}
\nabla\times\v A_K =\frac{\mu_0}{4\pi}\int d^3x'\frac{1}{R}[\nabla'\times\v J].
\end{equation}
Evidently, Eq.~(40) is identified with the usual retarded form of the magnetic field
\begin{equation}
\v B=\nabla\times\v A_K =\frac{\mu_0}{4\pi}\int d^3x'\frac{1}{R}[\nabla'\times\v J].
\end{equation}
Therefore, we do not require the complicated Eqs.~(32) to verify that the Kirchhoff-gauge potentials yield the retarded electric and magnetic fields. 

\section {VELOCITY GAUGE}
The velocity gauge (v-gauge) is one in which the scalar potential propagates with an arbitrary speed. This gauge is not very well known despite the fact that it was proposed by Yang\cite{12} several years ago. The v-gauge is really a family of gauges that contains the Lorenz and Coulomb gauges as particular members. It also includes the Kirchhoff gauge.\cite{9} The v-gauge has recently emphasized by Drury,\cite{19} Jackson\cite{5} and Yang\cite{11} himself. 
If we assume the v-gauge defined by Eq.~(7) then Eqs.~(3) become
\begin{subequations}
\begin{align}
\nabla^2\Phi_v- \frac{1}{v^2}\frac{\partial^2\Phi}{\partial t^2}&=\! -\frac{\rho}{\epsilon_0},\\
\Box^2\v A_v&=\!-\mu_0\v J + \frac{1}{c^2}\nabla\left(1-\frac{c^2}{v^2}\right)\frac
{\partial \Phi_v}{\partial t},
\end{align}
\end{subequations}
where $\Phi_v$ and $\v A_v$ denote the potentials are in the velocity gauge. The generality of Eqs.~(42) become evident when one observes that they reduce to Eqs.~(9) when $v=c$; to Eqs.~(19) when $v=\infty$ and to Eqs.~(30) when $v=ic$. The solutions of Eqs.~(42) are given by\cite{5}
\begin{subequations}
\begin{align}
\Phi_v(\v x,t)=& \frac{1}{4\pi\epsilon_0}\int d^3x' \frac{1}{R}\rho(\v x',t-R/v)),\\
\v A_v(\v x,t)=&\frac{\mu_0}{4\pi}\int d^3 x'\frac{1}{R}\bigg(\left\{\v J(\v x',t-R/c)-c\hat{\v R}\rho(\v x',t-R/c)\right\} +\frac{c^2}{v}\hat{\v R}\rho(\v x',t-R/v)\nonumber\\
\qquad \qquad &+\frac{c^2\hat{\v R}}{R} \int_{R/v}^{R/c}d\tau\;\rho(\v x',t-\tau)\bigg).
\end{align}
\end{subequations}
As expected, Eqs.~(43) reduces to Eqs.~(10) when $v=c$, to Eqs.~(20) when $v=\infty$ and to Eqs.~(32) when $v=ic$. 
According to Eq.~(42a), the potential $\Phi_v$ propagates with an arbitrary speed $v$ which may be subluminal $(v<c)$ or luminal $(v=c)$ or superluminal $(v>c)$ including the instantaneous limit $(v=\infty)$. In Ref.~5 it has been shown that Eqs.~(43) yield the retarded electric and magnetic fields. With regard to the v-gauge, Jackson has emphasized:\cite{5} ``The v-gauge illustrates dramatically how arbitrary and
unphysical the {\it potentials} can be, yet still yield the same physically sensible {\it fields}." The velocity gauge cannot be written in a relativistically covariant form. 

The v-gauge scalar potential $\Phi_v$ generates the field\cite{5} 
\begin{equation}
-\nabla \Phi_v(\v x,t)=\frac{1}{4\pi\epsilon_0}\int d^3 x'\left(\frac{\hat{\v R}}{R^2}\rho(\v x',t-R/v)+\frac{\hat{\v R}}{Rv}\frac{\partial\rho(\v x',t-R/v)}{\partial (t-R/v)} 
\right). 
\end{equation}
Evidently, this term does not displays the property of propagation at the speed of light $c$. The presence of the term $-\nabla \Phi_v$ possessing  an arbitrary propagation into the electric field expressed in terms of the v-gauge potentials: $\v E=-\nabla\Phi_v-\partial \v A_v/\partial t$ is also surprising at first sight. In particular, when $v$ is superluminal we have a conflict with special relativity which states that no physical information can propagate faster than $c$ in vacuum. However, at this stage the reader can reasonably suspect that, like the instantaneous term $-\nabla \Phi_C$ or the imaginary term $-\nabla \Phi_K$, the arbitrarily-propagated term  $-\nabla \Phi_v$ does not play a physical role in the electric field. 
To understand the role played by the term $-\nabla \Phi_v$ in the field $\v E=-\nabla\Phi_v-\partial \v A_v/\partial t$, let us apply the method proposed here to show that the potentials $\Phi_v$ and $\v A_v$ yield the fields $\v E$ and $\v B$.

Step 1. After applying the velocity condition (7) to Eqs.~(3), we obtained Eqs.~(42). We symmetrize Eq.~(42a) with respect to Eq.~(42b) by adding the term $-(1/c^2)\partial^2\Phi_v/\partial t^2$ on both sides of Eq.~(42a). The resulting equation can be written as 
\begin{equation}
\Box^2\Phi_v= -\frac{\rho }{\epsilon_0}-\frac{1}{c^2}\bigg(1-\frac{c^2}{v^2}\bigg)\frac
{\partial^2 \Phi_v}{\partial t^2}. 
\end{equation}

Step 2. We take minus the gradient to Eq.~(45) and minus the time derivative to Eq.~(42b). The resulting equations are
\begin{subequations}
\begin{align}
-\Box^2\nabla\Phi_v&= \frac{1}{\epsilon_0}\nabla\rho+\frac{1}{c^2}\nabla\left(1-\frac{c^2}{v^2}\right)\frac
{\partial^2 \Phi_v}{\partial t^2},\\
-\Box^2\frac{\partial \v A_v}{\partial t }&=\mu_0\frac{\partial \v J}{\partial t }-\frac{1}{c^2}\nabla\left(1-\frac{c^2}{v^2}\right)\frac
{\partial^2 \Phi_v}{\partial t^2}.
\end{align}
\end{subequations}

Step 3. We add Eqs.~(46) to obtain the wave equation 
\begin{equation}
\Box^2\left(- \nabla\Phi_v-\frac{\partial \v A_v}{\partial t }\right)= \frac{\nabla\rho}{\epsilon_0}+\mu_0\frac{\partial \v J}{\partial t },
\end{equation}
with the retarded solution 
\begin{equation}
-\frac{\partial \v A_v}{\partial t} =\frac{1}{4\pi\epsilon_0}\int d^3x'\frac{1}{R}\left[ -\nabla'\rho-\frac{1}{c^2}\frac{\partial \v J}{\partial t' }\right]+ \nabla\Phi_v.
\end{equation}
We observe that the term $-\partial \v A_v/\partial t$ in Eq.~(48) contains the component $\nabla\Phi_v$ which cancels exactly the term $- \nabla\Phi_v$ with an arbitrary propagation that appears in  the electric field $\v E=-\nabla\Phi_v-\partial \v A_v/\partial t.$ In other words, the explicit presence of a term possessing an arbitrary propagation $(-\nabla\Phi_v)$ in the electric field expressed in terms of the v-gauge potentials is irrelevant because such a term is always canceled by one of the components $(\nabla\Phi_v)$ of the v-gauge vector potential ($-\partial \v A_v/\partial t$). This means that the field $-\nabla\Phi_v$ is a $spurious$ field with mathematical but not physical meaning and that the propagation at the speed of light $c$ of the electric field is never effectively lost. Therefore, when Eq.~(48) is used into the expression  $\v E=-\nabla\Phi_v-\partial \v A_v/\partial t$ we obtain the usual retarded form of the electric field
\begin{equation}
\v E=-\nabla\Phi_v-\frac{\partial \v A_v}{\partial t}=\frac{1}{4\pi\epsilon_0}\int d^3x'\frac{1}{R}\left[ -\nabla'\rho-\frac{1}{c^2}\frac{\partial \v J}{\partial t' }\right],\\
\end{equation}

Step 4. We take the curl to Eq.~(42b) to obtain the wave equation 
\begin{equation}
\square^2\left(\nabla\times\v A_v\right)=-\mu_0\nabla\times\v J.
\end{equation}
with the retarded solution
\begin{equation}
\nabla\times\v A_v =\frac{\mu_0}{4\pi\epsilon_0}\int d^3x'\frac{1}{R}[\nabla'\times\v J].
\end{equation}
It is evident that Eq.~(51) gives the usual retarded form of the magnetic field
\begin{equation}
\v B=\nabla\times\v A_v =\frac{\mu_0}{4\pi\epsilon_0}\int d^3x'\frac{1}{R}[\nabla'\times\v J].
\end{equation}
Therefore, we do not require the complicated Eqs.~(43) to show that the potentials in the velocity gauge yield the retarded electric and magnetic fields.

\section {TEMPORAL GAUGE}
In previous sections we have pointed out that the scalar potential can be instantaneous, imaginary and (in particular) superluminal depending on the adopted gauge (Coulomb, Kirchhoff and velocity gauges respectively). Now we will see that the scalar potential can also not exist! In fact, the temporal gauge is one in which the scalar potential is identically zero.\cite{5} This means that the electric and magnetic fields are defined only by the vector potential: 
\begin{subequations}
\begin{align}
\v E&=\!-\frac{\partial \v A_T}{\partial t},\\
\v B&=\!\nabla\times \v A_T
\end{align}
\end{subequations}
where we have used the notation $\v A_{T}$ to specify that the vector potential is in the temporal gauge. It is evident that the temporal gauge cannot be written in a relativistically covariant form. We should mention that the temporal gauge is not usually introduced in textbooks of classical electrodynamics. Having in mind the fact that the scalar potential in the Lorenz, Coulomb, Kirchhoff and velocity gauges is always defined by the charge density, the reader might wonder why the scalar potential in the temporal gauge does not exist despite that there is a non zero charge density. The simple answer to this question is that the values of the charge density do not necessarily originate a scalar potential in all gauges. The existence of a scalar potential (with different propagation properties) generally depends on the adopted gauge. In other words, the retarded values of the charge density always physically contribute to the electric field but they do not originate a scalar potential in the temporal gauge. 

If we assume the temporal gauge defined by Eq.~(8) then Eqs.~(3) become
\begin{subequations}
\begin{align}
\frac{\partial}{\partial t}(\nabla\cdot \v A_{T})&= -\frac{\rho}{\epsilon_0},\\
\square^2\v A_{T}&=-\mu_0\v J + \nabla(\nabla\cdot\v A_{T}),
\end{align}
\end{subequations}
The solution of Eqs. (54) is given by\cite{5}
\begin{equation}
\v A_T(\v x,t)=\frac{\mu_0}{4\pi}\int d^3 x' \frac{1}{R}\left(\left\{\v J(\v x',t-R/c)-c\hat{\v R}\rho(\v x',t-R/c)\right\}-\frac{c^2\hat{\v R}}{R} \int_{R/c}^{t-t_0}d\tau\rho(\v x',t-\tau)\right).
\end{equation}
In Ref. 5 it has been demonstrated that the potential $\v A_T$ in Eq.~(55) yields the retarded electric and magnetic fields. In contrast to other gauges, in the temporal gauge both the charge density and the current density are related exclusively with the vector potential. With the idea of understanding this result, 
let us apply the proposed method in this paper to show that the temporal potential $\v A_T$ yields the retarded fields $\v E$ and $\v B.$ 

Step 1. After applying the temporal condition (8) to Eqs.~(3), we obtained Eqs.~(54). It is not necessary to symmetrize Eq.~(54a) with respect to Eq.~(54b). 

Step 2. We take minus the gradient to Eq.~(54a) and minus the time derivative to Eq.~(54b) to obtain the equations 
\begin{subequations}
\begin{align}
-\frac{\partial}{\partial t}\nabla(\nabla\cdot \v A_{T})&=\! \frac{1}{\epsilon_0}\nabla\rho,\\
-\square^2\frac{\partial \v A_T}{\partial t }&=\!\mu_0\frac{\partial \v J}{\partial t }-\frac{\partial}{\partial t} \nabla(\nabla\cdot\v A_{T}).
\end{align}
\end{subequations}

Step 3. We add Eqs.~(56) to obtain the wave equation 
\begin{equation}
\square^2\left(-\frac{\partial \v A_T}{\partial t }\right)= \frac{1}{\epsilon_0}\nabla\rho+\mu_0\frac{\partial \v J}{\partial t },
\end{equation}
with the retarded solution 
\begin{equation}
-\frac{\partial \v A_T}{\partial t} =\frac{1}{4\pi\epsilon_0}\int d^3x'\frac{1}{R}\Bigg[ -\nabla'\rho-\frac{1}{c^2}\frac{\partial \v J}{\partial t' }\Bigg].
\end{equation}
In contrast to the Coulomb, Kirchhoff and velocity gauges, in the temporal gauge we have not an additional unphysical term on the right-hand side of the expression for the time derivative of the vector potential. When Eq.~(58) is used in the expression $\v E=-\partial \v A_T/\partial t$, we obtain the usual retarded form of the electric field
\begin{equation}
\v E=-\frac{\partial \v A_T}{\partial t}=\frac{1}{4\pi\epsilon_0}\int d^3x'\frac{1}{R}\left[ -\nabla'\rho-\frac{1}{c^2}\frac{\partial \v J}{\partial t' }\right].
\end{equation}
 
Step 4. We take the curl to Eq.~(54b) to obtain the wave equation 
\begin{equation}
\square^2(\nabla\times \v A_T)=-\mu_0\nabla\times\v J.
\end{equation}
with the retarded solution
\begin{equation}
\nabla\times\v A_T =\frac{\mu_0}{4\pi}\int d^3x'\frac{1}{R}[\nabla'\times\v J].
\end{equation}
which leads directly to the usual retarded form of the magnetic field
\begin{equation}
\v B=\nabla\times\v A_T =\frac{\mu_0}{4\pi}\int d^3x'\frac{1}{R}[\nabla'\times\v J].
\end{equation}
The complicated Eq.~(55) is not required to show that the vector potential in the temporal gauge yields the retarded electric and magnetic fields.

\section {the non spurious character of $-\nabla\Phi_L$}
The fact that the Coulomb-gauge scalar potential $\Phi_C$ propagates instantaneously is not so worrying if we accept the general belief that 
potentials in electromagnetism are not physically measurable quantities. As pointed out by Griffiths:\cite{2} ``The point is that $V$ [the Coulomb-gauge scalar potential] {\it by itself} \;is not a physically measurable quantity." It follows that the instantaneous term $-\nabla\Phi_C$ [Eq.~(21)] must be also an unphysical quantity. But the subtle point is that $-\nabla\Phi_C$ is a part of the physical electric field expressed in terms of the Coulomb-gauge potentials:  $\v E=-\nabla\Phi_C- \partial \v A_C/\partial t$. We have pointed out that the presence of $-\nabla\Phi_C$ in the electric field is entirely irrelevant because it is always canceled by one component  of the remaining term $- \partial \v A_C/\partial t$. We have noted that 
$-\nabla\Phi_C$ is a purely formal result of the theory which is lacking of physical meaning. We have drawn similar conclusions for the term with imaginary propagation $-\nabla\Phi_K$ [Eq.~(33)] due to the Kirchhoff-gauge potential $\Phi_K$ and for the term with arbitrary propagation $-\nabla\Phi_v$ [Eq.~(44)] due to the v-gauge potential $\Phi_v$. Nevertheless, we cannot draw the same conclusion with respect to the term $-\nabla\Phi_L$ [Eq.~(11)] due to the Lorenz-gauge potential $\Phi_L$ because this term displays the experimentally verified properties of causality and propagation at the speed of light ---consistently the term $-\nabla\Phi_L$ is not canceled by some piece of the remaining term $- \partial \v A_L/\partial t$ of the electric field. 

Therefore, we can reasonably conclude that $-\nabla\Phi_L$ {\it is not} a spurious term like the terms $-\nabla\Phi_C,\;-\nabla\Phi_K$ and $-\nabla\Phi_v$. It follows that $-\nabla\Phi_L$ could be interpreted as a physical quantity in principle. Similarly, the term $- \partial \v A_L/\partial t$ due to the Lorenz-gauge potential also displays the properties of causality and propagation at the speed of light, which indicates that this term should not be interpreted as a spurious term. Thus, $- \partial \v A_L/\partial t$ could also be interpreted as a physical quantity. Moreover, the possible physical character of each one of the terms $-\nabla\Phi_L$ and $- \partial \v A_L/\partial t$ is strongly supported by the fact that the combination $-\nabla\Phi_L- \partial \v A_L/\partial t$, that is, the electric field, is physically detectable.\cite{20}

The remarkable fact is that the Lorenz-gauge potentials $\Phi_L$ and $\v A_L$ {\it naturally} inherit to the electric and magnetic fields the physical properties of causality and propagation at the speed of light. This result {\it suggests} that the Lorenz-gauge potentials (and not the Coulomb, Kirchhoff and velocity potentials) could be interpreted as physical quantities.

\section {concluding remarks}
Jackson has pointed out that:\cite{5} ``It seems necessary from time to time to show that the electric and magnetic $fields$ are independent of the choice of gauge for the potentials." Ordinarily, such a demonstration requires the previous derivation of explicit expressions for potentials. Nevertheless, for most gauges the derivation of potentials is generally simple for the scalar potential but generally difficult for the vector potential. 

In this paper we have demonstrated that the fields are independent of the choice of gauge for potentials in a variety of gauges (Lorentz, Coulomb, Kirchhoff, velocity and temporal gauges), that is, we have proposed a simple method, defined by the four steps appearing in Sec. V,  
that can be used to easily demonstrate that the potentials in these gauges yield the same retarded electric and magnetic fields.  Instead of using explicit expressions for the scalar and vector potentials in the above gauges, the method uses the dynamical equations of these potentials to obtain two wave equations, whose retarded solutions lead to the retarded fields. We have clearly identified the spurious character of the gradient of the scalar potential in the Coulomb, Kirchhoff and velocity gauges. We have emphasized the non spurious character of the scalar potential in the Lorenz gauge. Finally, we have suggested that the Lorenz-gauge potentials could be interpreted as physical quantities.

The method proposed in this paper is simple enough that it can be used in an advanced undergraduate course based on a text like Griffiths's book\cite{2} and in a graduate course based on a text like Jackson's book.\cite{1}

\vskip 9pt
\noindent {\bf Acknowledgments}
 
The author is grateful to an anonymous referee for his valuable comments and to Professor R. F. O'Connell for the kind hospitality extended to him in    
the Department of Physics and Astronomy of the Louisiana State University.


\begin{thebibliography}{99}

\bibitem{1} 
J. D. Jackson,  \textit{Classical Electrodynamics} (John Wiley, New York, 1999), 3rd ed. p. 242.

\bibitem{2} 
D. J. Griffiths,  {\it Introduction to Electrodynamics} (Prentice Hall, Englewood, NJ, 1999) 3rd ed., p. 421.

\bibitem{3} 
O. L. Brill and B. Goodman, ``Causality in the Coulomb gauge,"  Am. J. Phys. {\bf 35}, 832--837 (1967).

\bibitem{4}
J. D. Jackson and L. B. Okun, ``Historical roots of gauge invariance," Re. Mod. Phys. {\bf 73}, 663--680 (2001).

\bibitem{5} 
J. D. Jackson, ``From Lorenz to Coulomb and other explicit gauge transformations," Am. J. Phys. {\bf 70}, 917--928 (2002).

\bibitem{6} 
F. Rohrlich, ``Causality, the Coulomb field, and
Newton's law of gravitation," Am. J. Phys. {\bf 70},
411--414 (2002).

\bibitem{7} J. A. Heras, ``Comment on `Causality, the Coulomb field, and
Newton's law of gravitation' by F. Rohrlich [Am. J. Phys. {\bf 70},
411--414 (2002)]," Am. J. Phys. {\bf 71}, 729--730 (2003).

\bibitem{8} J. A. Heras, ``Instantaneous fields in classical
electrodynamics," Europhys. Lett. {\bf 69}, 1--7 (2005).

\bibitem{9} J. A. Heras, ``The Kirchhoff gauge," Ann. Phys. {\bf 321},
1265--1273 (2006).

\bibitem{10} G. Kirchhoff, ``II. Ueber die Bewegung der Elektricit${\ddot{\rm a}}$t in Leitern," 
Ann. Phys. Chem. {\bf 102}, 529–544 (1857). Reprinted in Gesammelte Abhandlungen von G. Kirchhoff (J. A. Barth, Leipzig, 1882), pp. 154–168.

\bibitem{11} K-H Yang, ``The physics of gauge transformations," Am. J. Phys. {\bf 73}, 742--751 (2005).

 \bibitem{12} K.-H. Yang, ``Gauge transformations and quantum mechanics. II. Physical interpretation of classical gauge transformations," Ann. Phys. (N.Y.) 101, 97–118 (1976).

\bibitem{13}
O. D. Jefimenko, {\it Electricity and Magnetism} (Electret
Scientific, Star City, WV, 1989), 2nd ed., p. 516. See also Ref. 1, p. 247.

\bibitem{14} J. A. Heras, ``Comment on `A generalized Helmholtz theorem for time-varying vector fields,'  by Artice. M. Davis [Am. J. Phys. {\bf 74}, 72-76 (2006)]," Am. J. Phys. {\bf 74}, 743--745 (2006).

\bibitem{15} J. A. Heras, ``Comment on `Helmholtz theorem and the v-gauge in the problem of superluminal and instantaneous signals in classical electrodynamics,' by Chubykalo et al [Found. Phys. Lett. {\bf 19}, 37-49 (2006)],"
Found. Phys. Lett. (to be published).


\bibitem{16} 
It should be noted that in the proposed method we could also {\it choose} the {\it advanced} solutions ---those evaluated at the advanced time $t'=t+R/c$--- of the wave equations derived in the steps 3 and 4.  This would lead to the advanced form of the electric and magnetic fields.

\bibitem{17} 
E. Zampino, ``A brief study of the transformations of Maxwell's equations in Euclidean four space," J. Math. Phys. {\bf  27}, 1315-1318 (1986).

\bibitem{18} 
J. A. Heras, ``Euclidean electromagnetism in four space: A discussion between God and the Devil," Am. J. Phys. {\bf 62} 914-916 (1994).
 
 \bibitem{19}
D. M. Drury, ``The unification of the Lorentz and Coulomb gauges of electromagnetic theory," IEEE Trans. Educ. {\bf 43} (1), 69-72 (2000).


\bibitem{20}
It can be argued, however, that from the fact that the combination of terms $-\nabla\Phi_L- \partial \v A_L/\partial t$ can be physically detected one cannot necessarily imply that each one of them could be detected separately.

 
\end{thebibliography}
\end{document}